\input{aipcheck}

\documentclass[
    ,final            
  ]
  {aipproc}

\layoutstyle{6x9}

\newcommand{\hst}{{\em HST}}
\newcommand{\vlt}{{\em VLT}}
\newcommand{\wfpc}{{\em WFPC2}}
\newcommand{\xmm}{{\em XMM-Newton}}
\newcommand{\chan}{{\em Chandra}}

\begin{document}

\title{Neutron Star Astronomy with the HST}


\author{Roberto P. Mignani}{address={Mullard Space Science Laboratory - University College London}}

\begin{abstract}
Since its launch  in 1990, \hst\ has played a  leading role in optical
studies  of  isolated  neutron  stars,  both  radio-loud  pulsars  and
radio-silent ones, paving the  way to follow-up observations performed
with 8m-class telescopes, like the \vlt, the Gemini, and Subaru. Here,
I present  the last results obtained  mostly by the  \wfpc, before its
de-commissioning during  the last  refurbishment mission in  May 2009,
from   the  observations   of  the   rotation-powered   pulsars  PSR\,
B0540$-$69, PSR\, B1055$-$52 and of  the CCO 1E\, 1207.4$-$5209 in the
PKS\, 1209$-$52 SNR.
\end{abstract}

\maketitle

\section{Introduction}

In its  20 years of activity,  \hst\ has observed  almost all isolated
neutron stars (INSs) with an optical counterpart, 24 to date including
possible  candidates (see, e.g.  Mignani 2010a),  and has  carried out
deep searches for several  others. In particular, \hst\ has discovered
the  optical counterpart  to seven  INSs. Four  of them  are classical
rotation-powered  pulsars,  one  of   which  is  the  first  ms-pulsar
identified in  the optical,  the others are  X-ray Dim  INSs (XDINSs).
Among the  major \hst\  achievement in optical  studies of  INSs (see,
e.g.  Mignani 2010b),  one  should  note the  many  detections in  the
near-IR  and in  the near-UV,  high spatial  resolution phase-averaged
polarimetry observations  of pulsars and  their associated pulsar-wind
nebulae   (PWNe),  proper   motion  and,   in  some   cases,  parallax
measurements for several rotation-powered pulsars and XDINSs.  Here, I
present the results of the  last observations of INSs performed by the
\wfpc.

\section{PSR\, B0540$-$69}

PSR\, B0540$-$69 is a young ($\sim 1600$ yr old) pulsar ($P=50$ ms) in
the  LMC, discovered in  the X-rays,  and soon  after observed  in the
optical with  a $V\sim  22.5$ counterpart (Caraveo  et al.  1992), and
only later  in radio. Because of  its similar energetic  ($\dot{E}
\sim 1.5 \times 10^{38}$ erg s$^{-1}$) and magnetic field ($B\sim 5 \times 10^{12}$ G) is also
referred as  the ``Crab twin''.  We observed PSR\, B0540$-$69  for six
\hst\ orbits:  four were devoted  to polarimetry observations  (1800 s
each) through the 606W+POLQ filter and two to multi-band photometry in
the 336W (780 s), 450W (780 s),  555W (300 s), 675W (420 s), 814W (420
s) filters,  to provide, for  the first time, an  homogeneous spectral
coverage from $\sim$ 3000 to 8000 \AA.
First of all, we used  our imaging observations to set new constraints
on the  pulsar proper  motion. Using a  sample of  \wfpc\ observations
spanning  4 years,  Serafimovich et  al. (2004)  claimed  a marginally
significant  proper motion of  $4.9 \pm  2.3$ mas  yr$^{-1}$, possibly
parallel to the  nebula symmetry axis, as in the case  of the Crab and
Vela pulsar/PWNe.   However, more recent  \wfpc\ observations spanning
10 years (De Luca et al. 2007), did not confirm this claim and yielded
a proper motion upper limit of 1.2 mas yr$^{-1}$ ($3 \sigma$). Our new
\wfpc\ observations, extending the  time baseline to 12 years, allowed
to refine the upper limit to  1 mas yr$^{-1}$ ($3 \sigma$), implying a
transverse velocity  $v_t < 250$  km s$^{-1}$(Mignani et  al.  2010a).
At the same time, we  used the available multi-epoch image database to
compute    the   pulsar    coordinates   through    absolute   optical
astrometry. From the average  of different measurements, we determined
the pulsar position with an accuracy of 5 mas per coordinate ($1
\sigma$), which supersedes in accuracy previously reported values (see
Mignani et al. 2010a) and has been  used by Gradari et al. (2010) as a
reference for the pulsar optical timing observations.
We used our multi-band imaging observations, together with archival
\wfpc\  broad band  images,  to accurately  characterise the  pulsar's
optical  spectrum.   Middleditch  et  al.  (1987)  showed  a  possible
evidence of a flux excess in the U-band with respect to the underlying
power-law  (PL) continuum,  not confirmed  by Nasuti  et  al.  (1997).
More recently, using a small set of broad/medium-band
\wfpc\ images, Serafimovich et al.  (2004) measured a PL with spectral
index $\alpha = 1.07 \pm  0.2$. Using our extended \wfpc\ database, we
measured  $\alpha= 0.70  \pm 0.07$,  more consistent  with  the values
measured in other rotation-powered pulsars, with no evidence of a flux
excess in the  U-band. Interestingly enough, the optical  PL lies well
below the extrapolation of the X-ray one, which implies a double break
in  the optical--to--X-ray spectrum,  never observed  so far  in other
rotation-powered  pulsars.    For  PSR\,  B0540$-$69,   only  a  \vlt\
meausurement  of a  phase-averaged polarisation  degree (PD)  of $\sim
5$\% (quoted without errors) existed prior to our observations (Wagner
\& Seifert 2000), probably polluted by the contribution of the SNR and
by  the  difficulties  in  resolving  the pulsar  in  the  low-spatial
resolution \vlt\  images.  After  subtracting the contribution  of the
foreground  polarisation  and  of   the  SNR,  we  measured  a  pulsar
polarisation   $PD=16\%\pm4\%$,  with   a  position   angle   (PA)  of
$22\pm12^{\circ}$.  This  is closely aligned with  the possible proper
motion direction of the knot observed southwest of the pulsar (De Luca
et al.  2007),  which is also polarised ($PD \sim  10$\%), with its PA
approximately aligned with that of the pulsar.

 \section{PSR\, B1055$-$52}

 PSR\, B1055$-$52 is  a middle-aged ($\sim$ 535 kyr  old) radio, X and
 $\gamma$-ray pulsar  ($\dot{E} \sim  3 \times 10^{34}$  erg s$^{-1}$;
 $B\sim 1.1 \times  10^{12}$ G) and, together with  PSR\, B0656+14 and
 Geminga, one of the  so-called ``Three Musketeers''.  In the optical,
 the  detection of  PSR\, B1055$-$52  has  been troublesome  due to  a
 bright  star ($V \sim  14$), located  at $\sim  4.4$ arcsec  from the
 pulsar, which  hampered all ground-based  observations (e.g., Mignani
 et  al. 2010b).   Indeed, the  pulsar counterpart  was  only detected
 thanks to the high spatial  resolution and near-UV sensitivity of the
 \hst/{\em FOC} (Mignani et  al.  1997).  We observed PSR\, B1055$-$52
 with the \hst\  both in the near-UV (140 LP filter;  5600 s) with the
 {\em ACS/SBC}  and in the optical  with the \wfpc\ in  the 450W (1800
 s),   555W  (1800s)   and  702W   (3600  s)   filters.    The  pulsar
 identification was complicated by  the uncertainty on its coordinates
 (epoch 1978  in the  ATNF catalogue)  which could be  up to  $\sim 4$
 arcseconds due  to its unknown  proper motion.  We detected  a source
 positionally   concident,  within  the   uncertainty  of   the  \hst\
 astrometry, with  the pulsar candidate  counterpart both in  the {\em
 SBC}  ($m_{140LP} \sim 22.6$),  and in  the \wfpc\  images ($m_{555W}
 \sim 25.4$; $m_{702W} \sim 26.08$; $m_{450W}>24.97$).  Thus, its blue
 colours  certified its  identification  with the  pulsar (Mignani  et
 al. 2010c).
Using  our  high-spatial  resolution  \wfpc\ images,  we  obtained  an
updated measurement of the  pulsar coordinates (epoch 2008.18) through
absolute optical astrometry  ($\sim 0.15$ arcsec positional accuracy),
consistent  with  the new  radio  timing  position (Manchester,  R.N.,
private  communication).  From  the  comparison with  the  ATNF  radio
coordinates  we also  obtained  the first  measurement  of the  pulsar
absolute  proper motion  which  is consistent  with  the relative  one
measured from  the comparison of  the pulsar position measured  in the
{\em SBC} and {\em FOC}  images, taken 12 years apart: $\mu_{\alpha} =
43  \pm  6$   mas  yr$^{-1}$  and  $\mu_{\delta}  =   -5  \pm  6$  mas
yr$^{-1}$. This implies a transverse velocity $v_t \sim (70
\pm 8)$  km s$^{-1}$, for a 
pulsar distance of 350 pc  (see below).  We then computed the backward
extrapolation of the  pulsar galactic orbit to locate  its birth place
and indentify its  parental open cluster, after backward-extrapolating
the orbits  of selected  candidate clusters.  Unfortunately,  both the
unknown  pulsar radial  velocity and  the uncertain  distance  did not
allow  us to determine  an unambiguous  association.  A  more accurate
determination of the pulsar distance, e.g. from the measurement of the
radio  parallax, will  be important  to narrow  the list  of candidate
parental clusters.
We used our multi-band images to characterise, for the first time, the
pulsar optical-UV spectrum.  This  can be described by the combination
of  a   power-law  (PL$_{\rm O}$;  $\alpha_{\rm O}   =  1.05  \pm  0.34$)   and  a
Rayleigh-Jeans  (RJ).  Interestingly  enough,  the  RJ  is  above  the
extrapolation of the cold blackbody (BB$_{\rm C}$) component used to fit the
\xmm\ spectrum  (De Luca et al.  2005), together with  a hot blackbody
(BB$_{\rm  H}$)  and a  power-law  (PL$_{\rm  X}$).   This suggests  a
three-component thermal  map for the neutron star  surface, unlike the
other ``Musketeers''.   The temperature of the  RJ component, however,
is  parametrised by the  ratio $(d/R)^2$,  where $d$  and $R$  are the
neutron star distance  and the radius of the  optical emitting region,
respectively.  To solve the degeneracy, we imposed that $R<R_{NS,13}$,
where $R_{NS,13}$  is the neutron  star radius in  units of 13  km (as
seen from infinity) and that the  BB extrapolation of the RJ yields an
X-ray flux  much smaller  than that of  BB$_{\rm C}$.  From  the limit
case $R\sim  R_{NS,13}$ we then  derive a distance estimate  of $d\sim
350$  pc,  i.e.   lower than  the  750  pc  value estimated  from  the
dispersion measure (DM). The new distance yields a factor of 4 smaller
radius (5.7 km) for BB$_{\rm C}$, which has important implications for
comparisons with  neutron star cooling  model, and implies  a downward
rescaling  for the  different multi-wavelength  emission efficiencies,
with the $\gamma$-ray one being  $\sim 0.13\%$.  The best fit PL$_{\rm
O}$ is below  the extrapolation in the optical  domain of PL${\rm _X}$
($\alpha_{\rm   X}  =   0.7  \pm   0.1$),  as   seen  in   many  other
rotation-powered pulsars.  This suggests  a spectral break between the
X-ray  and  optical-UV regions.   However,  the  extrapolation of  the
$\gamma$-ray  PL  ($\alpha_{\gamma} \sim  0.5$)  possibly matches  the
optical one,  as seen, e.g  in the Vela  pulsar, although it is  not a
common feature of all rotation-powered pulsars.


\section{The CCO in PKS\,1209$-$52}

The X-ray source 1E\, 1207.4$-$5209 in the PKS\, 1209$-$52 SNR is one of
the  three CCOs  found to feature X-ray  pulsations  so  far,  with a  period
$P=424$ms.  
Optical/IR  observations with  the
\hst/ACS and with the \vlt/ISAAC spotted a possible counterpart from a
claimed coincidence  with the source \chan\  position (Pavlov et
al.  2004).   
However, more recent astrometry recalibration of the \hst\ images 
 suggested that  the proposed  counterpart was  off the
\chan\ error  circle. This has  been confirmed also by  updated \chan\
astrometry  of the source  (see De  Luca et  al.  2011  and references
therein).   While  this  is  already  a strong  argument  against  the
association of the object with  the CCO, the final evidence comes from
the measurement of its proper motion with respect to the centre of the
SNR.  Indeed,  if associated  with  the  CCO,  the counterpart  should
feature a proper motion $\mu \sim 70$ mas yr$^{-1}$, given the SNR age
of $\sim$ 7000 years and the offset of the CCO with respect to the SNR
centre ($\sim 8$ arcmin).  We observed the CCO field with the
\wfpc\  in  the  814W  filter  (2000  s)  and  we  performed  relative
astrometry with  respect to archival {\em  ACS} observations performed
in 2003 through the same filter. We obtained a $3 \sigma$ limit on the
proper motion of $\mu = 7$  mas yr$^{-1}$, which would imply a SNR age
larger  than   70\,000  years.    Thus,  such  a   measurement  firmly
establishes that  the proposed counterpart is not  associated with the
CCO.  From  the available \hst\ and \vlt\  photometry, complemented by
public {\em  Spitzer} data, we could  set a limit of  $\sim 0.1$ solar
masses  on  a  stellar  mass  companion.   The  upper  limits  on  the
optical/IR spectrum are also compatible  with the presence of a debris
disc. Using disc emission models  (Perna et al.  2000) and the derived
CCO X-ray luminosity $L_X \sim  2.2 \times  10^{33}$  erg s$^{-1}$ at a 2.2 kpc distance we determined,  for  the assumed  disc
parameters (inclination, albedo, inner  radius) and for the constraint
on the magnetic  field ($B<8 \times 10^{10}$ G), an  upper limit of $8
\times 10^{11}$  g s$^{-1}$ on the  disc $\dot{M}$. This  is too small
for  a putative  disc to  significantly  contribute to  the polar  cap
re-heating, as it has been proposed in some CCO scenarios.


%
 
\section{Conclusions}

INS astronomy with  the refurbished \hst\ (up to  2015) has still many
potentials, thanks to its instrument suite, covering the near-UV (COS,
ACS/SBC, STIS), near-IR (NICMOS, WFC3) and optical (ACS/WFC, WFC3), as
well   as  timing   (COS,  STIS),   polarimetry  (ACS,   NICMOS),  and
spectroscopy capabilities. This allows  to study the NS magnetosphere,
the  NS  interior, to  search  for debris  discs,  and  to pursue  new
INS identifications.






\bibliographystyle{aipproc}   

\bibliography{sample}

\IfFileExists{\jobname.bbl}{}
 {\typeout{}
  \typeout{******************************************}
  \typeout{** Please run "bibtex \jobname" to optain}
  \typeout{** the bibliography and then re-run LaTeX}
  \typeout{** twice to fix the references!}
  \typeout{******************************************}
  \typeout{}
 }


\end{document}